\documentclass[aps,pre,twocolumn,groupedaddress,superscriptaddress,showpacs]{revtex4-1}
\usepackage{epsfig,amssymb,amsmath,graphicx,subfigure,hyperref}
\usepackage{tabularx}
\usepackage{float}
\usepackage{setspace}
\usepackage{color}
\usepackage{times}
\usepackage{verbatim}

\usepackage{array} 
\newcommand{\PreserveBackslash}[1]{\let\temp=\\#1\let\\=\temp} \newcolumntype{C}[1]{>{\PreserveBackslash\centering}p{#1}} \newcolumntype{R}[1]{>{\PreserveBackslash\raggedleft}p{#1}} \newcolumntype{L}[1]{>{\PreserveBackslash\raggedright}p{#1}} 
\usepackage{textcomp}

\begin{document}

\title{Randomness in self-assembled colloidal crystals can widen photonic band gaps through particle shape and internal structure}

\author{Duanduan Wan}
\affiliation{Department of Chemical Engineering, University of Michigan, Ann Arbor, Michigan 48109, USA}
\affiliation{School of Physics and Technology, Wuhan University, Wuhan 430072, China}
\author{Sharon C. Glotzer}
\email[E-mail:]{sglotzer@umich.edu}
\affiliation{Department of Chemical Engineering, University of Michigan, Ann Arbor, Michigan 48109, USA}
\affiliation{Department of Physics, University of Michigan, Ann Arbor, Michigan 48109, USA}
\affiliation{Department of Materials Science and Engineering and Biointerfaces Institute, University of Michigan, Ann Arbor, Michigan 48109, USA}

\date{\today}
\begin{abstract}
Using computer simulations, we explore how thermal noise-induced randomness in a self-assembled photonic crystal affects its photonic band gaps (PBGs). We consider a two-dimensional photonic crystal comprised of a self-assembled array of parallel dielectric hard rods of infinite length with circular or square cross section. We find the PBGs can exist over a large range of intermediate packing densities. Counterintuitively, the largest band gap does not always appear at the packing density where the crystal is most ordered, despite the randomness inherent in any self-assembled structure. For rods with square cross section at intermediate packing densities, we find that the transverse magnetic (TM) band gap of the self-assembled (i.e.~thermal) system can be larger than that of identical rods arranged in a perfect square lattice. By considering hollow rods, we find the band gap of transverse electric (TE) modes can be substantially increased while that of TM modes show no obvious improvement over solid rods. Our study suggests that particle shape and internal structure can be used to engineer the PBG of a self-assembled system despite the positional and orientational randomness arising from thermal noise.    
\end{abstract}
\maketitle

\section{Introduction}
An intriguing feature of colloids is their ability to self-assemble into ordered structures with interparticle distances commensurate with wavelengths of light \cite{Manoharan2015}.  Advances in  synthesis have produced a wide variety of anisootropic particles, such as polyhedra \cite{Henzie2012,Young2013, Gong2017}, dumbbells \cite{Forster2011}, spherocylinders \cite{Hosein2010}, superballs \cite{Meijer2017} and octapods \cite{Miszta2011}. Experiments and simulations have demonstrated a diverse range of close-packed superlattices whose structure depends on particle shape. Simulations of hard colloids (e.g., Refs.~\cite{Gang2011, Haji-Akbari2009, Kraft2012, Agarwal2011, Damasceno2012_science, Ni2012, Marechal2010, Chen2014, Wan2019}) predict complex crystals from an even larger variety of anisotropic shapes, which are versatile in terms of modification and functionalization.  Colloidal self-assembly is one route that has been explored to fabricate photonic crystals with photonic band gaps (PBGs) (e.g., Refs.~\cite{Ye2001,Vlasov2001, Hosein2010, Forster2011,Sowade2016}). Different from top-down design, this bottom-up method has advantages such as low cost and low energy consumption, and crystals can be produced over large areas \cite{Moon2010, Kim2011, Zhao2014}. Moreover, the particle size can be varied from tens of nanometers to micrometers in experiments, tuning the periodicity of crystals and consequently the PBG frequency \cite{Moon2010}. Besides the photonic crystals already realized in experiments, many other promising lattice structures from self-assembly approaches have been theoretically proposed (e.g., Refs.~\cite{Ding2014, Hynninen2007, Woldering2011, Pattabhiraman2017, Wang2017, Changizrezaei2017, Cersonsky2018, Lei2018, Cersonsky2019}). Previous studies of self-assembled photonic crystals have focused primarily on structure design and self-assembly pathways. How ``randomness" arising from the inevitable thermal noise in a self-assembled colloidal crystal affects PBGs has received little attention. At first glance one might expect noise to weaken or destroy photonic band gaps, but is this generally true?  If it is, it would suggest that thermodynamic self-assembly may not be a viable synthesis route for PBG materials.

Here we study the PBGs of self-assembled two-dimensional photonic colloidal crystals (Fig.~\ref{sa} shows some examples) using computer simulations. We consider self-assembled lattices of parallel dielectric rods of infinitely long length that interact through a hard core potential. We study rods of circular and square cross section and investigate a wide range of packing densities $\phi$ and dielectric constants 
$\epsilon$. Surprisingly, we find that the widest PBG does not always appear at the packing fraction where the rods are perfectly ordered. Moreover, for rods with square cross section at intermediate packing densities, we find that the transverse magnetic (TM) band gap of the self-assembled system can be larger than that of its corresponding perfect system. Further, we show that by considering hollow rods and optimizing the internal radius, the PBG of the transverse electric (TE) mode can be substantially increased while that of the TM mode does not show obvious improvement. We discuss the possibility of engineering PBGs in self-assembled colloidal systems by controlling relevant factors.   

\section{Methods}
To be consistent with previous work, we adapt the method used in Ref.~\cite{Michael2016} to generate self-assembled structures of cross sections of $N=200$ circular hard rods. For particles of other shapes, we performed Monte Carlo (MC) simulations with periodic boundary conditions using the hard particle Monte Carlo (HPMC) module in HOOMD-blue \cite{hoomd,Glaser2015}. Simulations were initialized at very low packing density ($\phi=0.01$) in a random configuration and slowly compressed to a target packing density using MC simulations. Here one MC step consists of $N+1$ trial moves including translation (plus rotation for rods of square cross section) of $N$ particles or rescaling of the box, where during each compression step the length of the sides of the box are rescaled to 0.9995 of their current value. At high packing densities, box rescaling can create  unphysical overlaps that allow us to reach higher densities; these overlaps are subsequently eliminated with isochoric Monte Carlo \cite{Haji-Akbari2009}. After the system reaches the targeted packing density, it is further equilibrated for $10^6$  MC steps. We use the supercell method \cite{Joannopoulos2008} implemented in the open source code MIT Photonic-Bands \cite{Johnson2001} to obtain the photonic band structure of equilibrated snapshots. All band gap sizes of self-assembled structures presented in this work are averaged over five independent runs.

\section{Results and discussion}

\begin{figure}
\centering 
\includegraphics[width=3.5in]{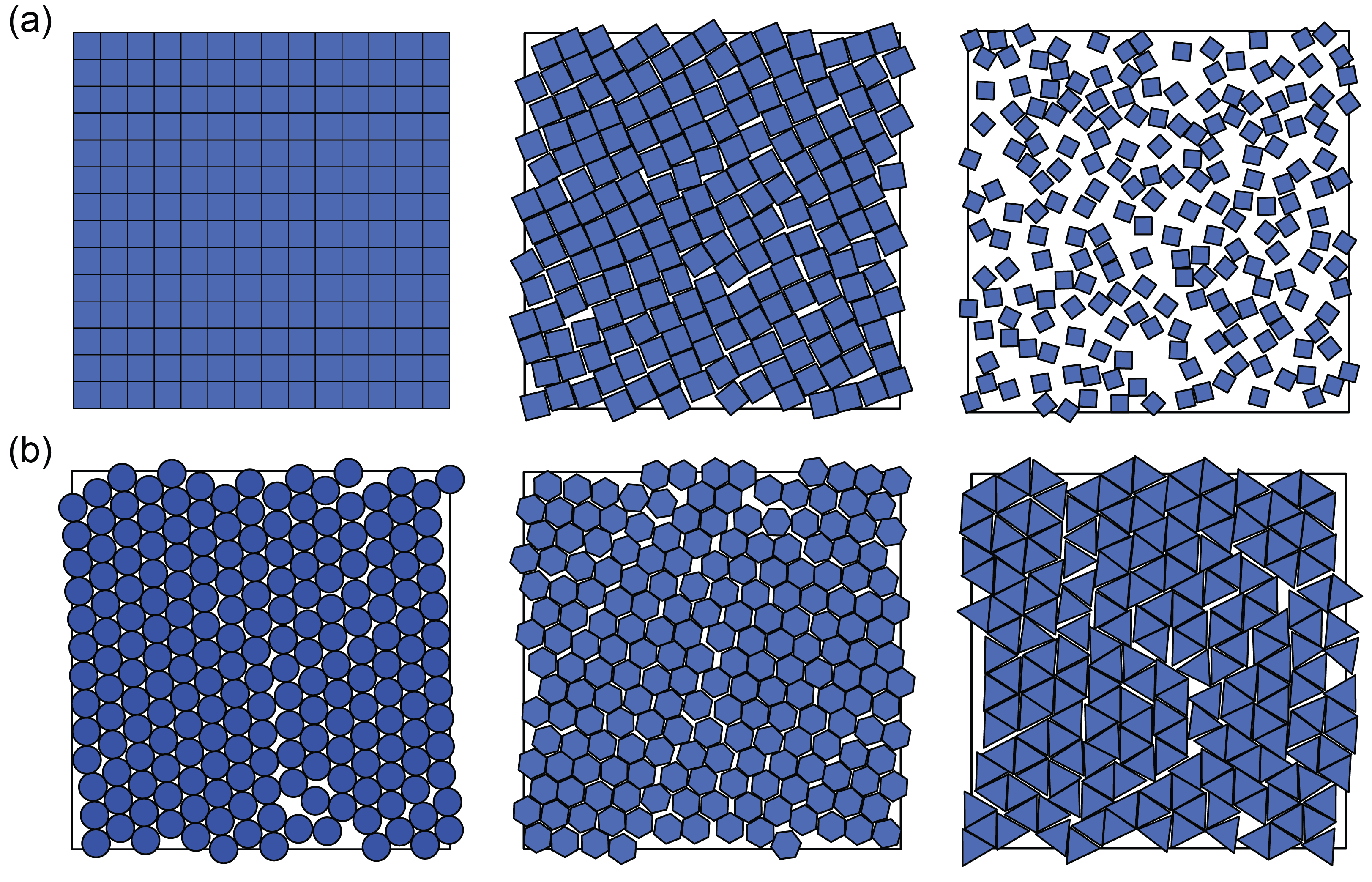} 
\caption{(Color online) Snapshots of cross section of a system of $N=200$ self-assembled rods described in the main text. (a) Rods of square cross section at packing density $\phi=1$ (homogeneous material, 196 rods shown), 0.85 and 0.4 (left to right).  (b) At $\phi=0.85$, equilibrium configurations of rods of circular, hexagonal and triangular cross sections (left to right).}
\label{sa}
\end{figure}

\begin{figure}
\centering 
\includegraphics[width=3.5in]{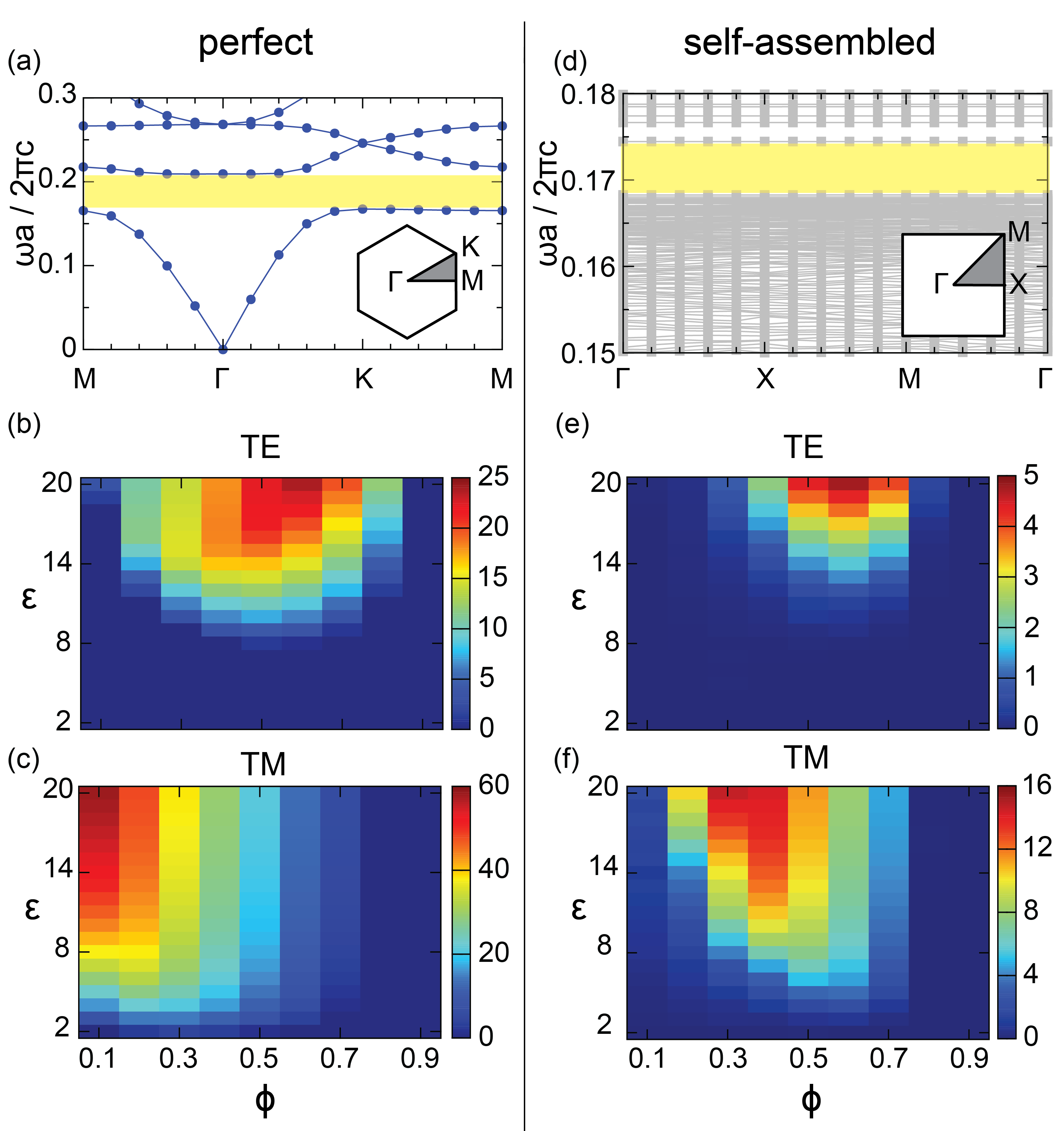} 
\caption{(Color online) Photonic band structure and band gap for arrays of parallel dielectric rods of circular cross section with radius $r=0.5a$ ($a$ is the diameter and the length unit) embedded in air ($\epsilon=1$). (a-c) Rods in a perfect triangular lattice. (a) An example of the band structure with TE polarization, at $\phi=0.5$ and $\epsilon=20$. The inset shows the first Brillouin zone for the periodic structure studied, with the symmetry points indicated. The yellow area represents the band gap observed. (b) Gap size (the percentage of $\triangle \omega / \omega_{0}$) as a function of $\phi$ and $\epsilon$ for the TE mode. (c) Gap size for the TM mode. (d-f) Parallel results for snapshots of $N=200$ self-assembled rods. Parameters in (d) are the same as those in (a). The band gap (e-f) is obtained by averaging over five independent simulation snapshots. } 
\label{disks}
\end{figure}

\begin{figure}
\centering 
\includegraphics[width=3.5in]{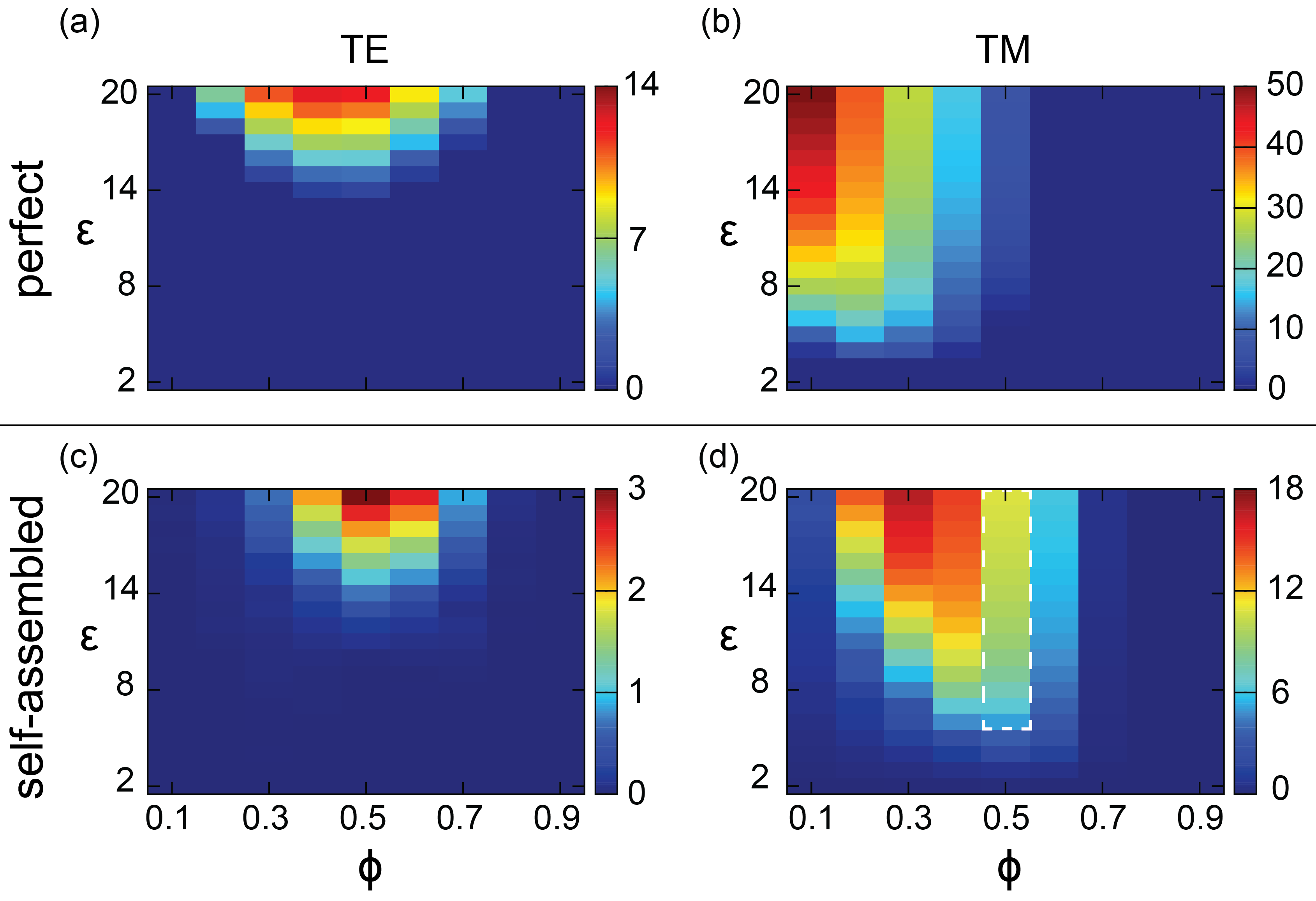} 
\caption{(Color online) Gap size as a function of $\phi$ and $\epsilon$ for rods with square cross section in a perfect square lattice (a-b) and the self-assembled crystal (c-d). The white dashed line in (d) indicates the region where both the perfect square lattice and the self-assembled crystal have a non-vanishing TM gap, and the gap size of the self-assembled structure is wider than in the perfect structure. }   
\label{squares}
\end{figure}

\begin{figure}
\centering 
\includegraphics[width=3.3in]{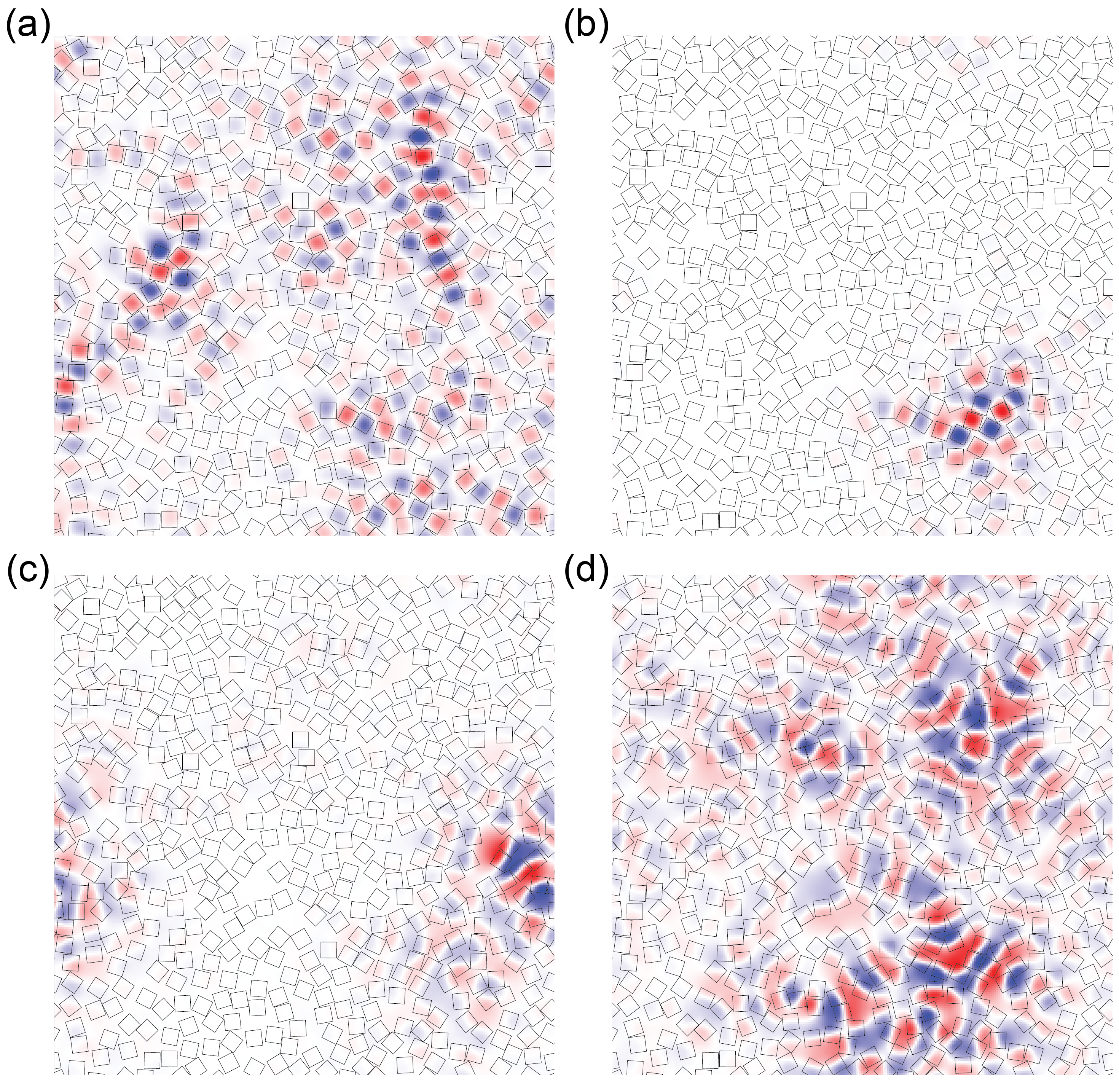} 
\caption{(Color online)  Electric field distribution in the system of $N=500$ self-assembled rods with square cross section for TM polarization: $\phi=0.5$ and $\epsilon=20$. (a) Extended mode before the PBG. (b) Localized mode before the PBG. (c) Localized mode after the PBG. (d) Extended mode after the PBG.}   
\label{modes}
\end{figure}

\begin{figure*}
\centering 
\includegraphics[width=7in]{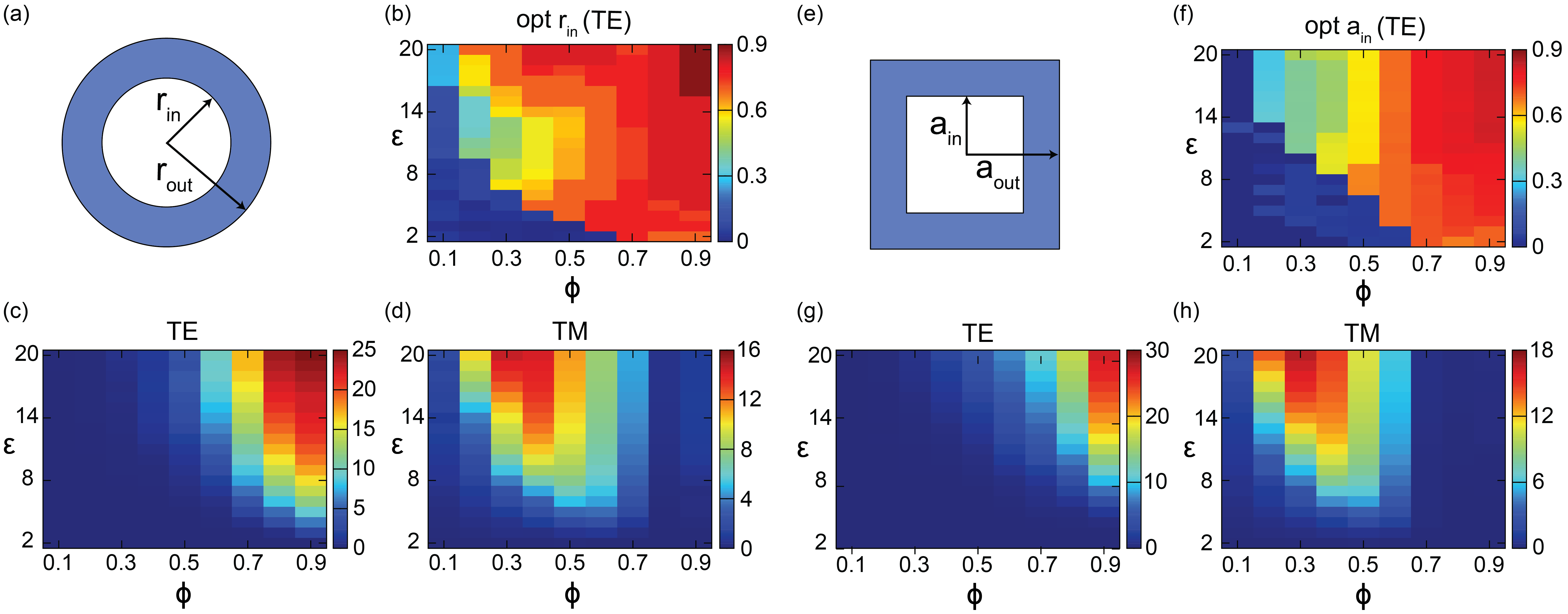} 
\caption{(Color online) Band gap for an array of parallel dielectric rods of annular cross section (a-d) or hollow square cross section (e-h). (a) Annular cross section. (b) Band structure at optimal $r_{in}$ value (described in the text) for TE polarization. Color indicates the value of $r_{in}/r_{out}$. (c-d) Band gap at optimal $r_{in}$, as a function of $\phi$ and $\epsilon$, for the self-assembled structure. (e) Hollow square cross section. (f) The optimal $a_{in}$ value for TE polarization. Color indicates the value of $a_{in}/a_{out}$. (g-h) Band gap at optimal $a_{in}$, as a function of $\phi$ and $\epsilon$, for the self-assembled structure.} 
\label{hollow}
\end{figure*}

We start from the photonic band structure for electromagnetic waves in a periodic array of parallel dielectric rods of circular cross section, whose intersection with a perpendicular plane form a perfect triangular lattice. This system was first studied in Ref.~\cite{Plihal1991}. The band structure can be calculated from a unit cell and Fig.~\ref{disks}(a) shows an example for TE polarization at $\phi=0.5$ and dielectric constant $\epsilon=20$. The yellow area indicates there is a band gap between the first and second bands. The relative gap size is defined as $\triangle \omega / \omega_{0}$, where $\triangle \omega$ is the width of the PBG and $\omega_{0}$ is the central frequency. Fig.~\ref{disks}(b and c) show the relative gap size as a function of $\phi$ and $\epsilon$. We investigated a wide range of $\phi$  and with $\epsilon$ in the range from 2 to 20.  Some low $\epsilon$  materials such as polystyrene and silica, and some high $\epsilon$ materials such as titania, selenium and amorphous silicon, fall in this range \cite{Moon2010}. For the TE mode, the largest PBG is at $\phi=0.6$ and $\epsilon=20$; for the TM mode, the largest PBG is at much lower packing density, i.e., $\phi=0.1$ and $\epsilon=20$. Furthermore, the largest band gap for the TM mode is much wider than that of the TE mode.
Fig.~\ref{disks}(d-f) show the equivalent results calculated for simulation snapshots of a self-assembled system of 200 rods (Fig.~\ref{sa}(b) (left)).  
Fig.~\ref{disks}(d) shows the band structure with the same parameters as those in Fig.~\ref{disks}(a). The yellow area indicates the band gap between the $N$th and the $(N+1)$th band. The band structure shows that the central frequency remains nearly constant, but the gap size decreases. Fig.~\ref{disks}(e,f) show the relative PBG over the entire range of parameters.  Because the bands are dense near the band edge and the deviation is relatively small, a sample size of five independent configurations is sufficient to obtain good statistics (see the Supplimentary Material for more information). We see that for the TE mode, the maximal band gap of the self-assembled system (Fig.~\ref{disks}(e)) appears at about $\phi=0.6$, similar to that in the perfect triangular system (Fig.~\ref{disks}(b)). At $\phi=0.9$, where the system is most ordered (close to the densest packing fraction 0.907), there is no band gap. We observe that for the TM mode, the packing fraction at which the maximal band gap appears shifts from $\phi=0.1$ (Fig.~\ref{disks}(c)) to $\phi=0.3$ (Fig.~\ref{disks}(f)), and there is no obvious band gap at $\phi=0.1$ in the self-assembled system. This is understandable because at very low density the system is highly disordered, where we do not expect a PBG. Thus, surprisingly, the largest gap does not occur when  the system is the most ordered. However, at the same $\phi$ and $\epsilon$ values, the self-assembled system always has a smaller PBG compared to that of the perfect ordered system. This is consistent with Ref.~\cite{Michael2016} that used a hard disk system to generate a seed pattern and placed cylindrical rods with an arbitrary fixed radius at the points of the seed pattern. Here we consider the seed pattern as the self-assembled structure of infinitly long rods. In practice, when rods are long enough and at high packing densities, they will tend to align parallel to each other \cite{Anders2014PNAS, Anders2014, Manoharan2015, Wan2018}.  Experiments have been able to obtain monodomain films of highly aligned carbon nanotubes from suspension [43]. We expect that similarly, by taking advantage of entropic effects and using auxiliary experimental skills, it is possible to align long colloidal rods as well.


Particle shape affects the PBG of a self-assembled crystal in two ways: (1) it defines the region of dielectric materials and (2) it determines the assembled structure and densest packing structure \cite{Bernard2011, Engel2013, Anderson2017}. Some examples are shown in Fig.~\ref{sa}. As a simple example, although infinite rods of circular or hexagonal cross section crystallize into triangular lattices, they have different maximum packing densities. Rods with square and triangular cross sections crystallize into square and hexagonal lattices, respectively, already below their maximum packing densities of 1.

To investigate the effect of cross-section shape on PGBs, we studied rods of square cross section (Fig.~\ref{squares}). The photonic band structure of rods with square cross section are similar to those for rods with circular cross section, but the gap widths, are generally different at the same packing density and with the same dielectric constant, which suggests the possibility of engineering PBGs using particle shape \cite{Cersonsky2018}. Moreover, the existence of PBGs in the square rod system suggests that Voronoi particles, which also tile space at densest packing, can be candidates for three-dimensional photonic crystals \cite{Schultz2015}. Another interesting aspect of the square rod system is that at some intermediate packing densities, the TM band gap of the self-assembled system can be larger than that of rods in a perfect square lattice at the same value of $\epsilon$. The white dashed line in Fig.~\ref{squares}(d) indicates the region ($\phi=0.5$, $\epsilon \in [6,20]$) where both the self-assembled structure and the perfect square lattice have a non-vanishing PBG, and the gap size of the former is wider. For example, at $\phi=0.5$ and $\epsilon=20$, the gap size of the self-assembled system is about $10.72\%$ and that of the perfect system is about $6.94\%$.  To further demonstrate the TM band gap, we show the electric field distribution of some TM modes around the PBG edges in Fig.~\ref{modes}. Localized and extended modes around a PBG gap have also been observed in other 2D systems (e.g.~\cite{Florescu2009, Ricouvier2019}). 
At $\phi=0.6$ and $\epsilon=20$, the gap size of the self-assembled system decreases to about $5.96\%$ (Fig.~\ref{squares}(d)) while that of the perfect system vanishes (at $\phi=0.6$, the perfect square lattice has no TM band gap for all $\epsilon$ values) (Fig.~\ref{squares}(b)).  Randomness due to fabrication errors in traditional lithography-based approaches is usually regarded as a bad aspect which deminishes PBGs, here instead we find a counterintuitive example where randomness actually helps to increase the PBG.

The PBGs for rods with circular or square cross section, however, are not large. In an attempt to increase the PBGs, we consider the design of the internal structure of the rods. Hollow or double-layer rods of various shapes and on various perfect lattice structures have been explored in previous studies \cite{Chang2006, Duque2012, Niu2013, Liu2013, Liu2015, Rezaei2006, Xiao2008}. Here we consider the simplest situation of hollow rods as indicated in Fig.~\ref{hollow}(a). We optimize the value of the inside radius $r_{in}$ to that which maximizes the TE(TM) band gap when the rods are in a perfect triangular lattice; the optimal $r_{in}$ for the TE mode is shown in Fig.~\ref{hollow}(b). Fig.~\ref{hollow}(c,d) show the band gap of the self-assembled structure at the optimal radius for the TE and TM modes, respectively. Compared with solid rods Fig.~\ref{disks}(e,f), the band gap of the TE mode is largely increased while that of the TM mode remains nearly the same. The value of $\phi$ at which the maximal PBG of the TE mode occurs changes from $\phi=0.6$ to $\phi=0.9$. Note that the band gap of the self-assembled structure with the optimal $r_{in}$ is smaller than that of the perfect lattice with the optimal $r_{in}$ (see the supplementary material). For rods with square cross section (Fig.~\ref{hollow}(e)), making the rods hollow also increases the PBGs of the TE mode while hardly changing the TM mode (Fig.~\ref{hollow}(f-h)). The optimal value of the inside side length $a_{in}$ maximizes the band gap when the rods are in a perfect square lattice.

\section{Conclusions}
We studied the PBGs of a self-assembled system consisting of parallel dielectric colloidal rods with circular or square cross section and interacting through a hard core potential. For the square rod system, we found that although the system has no PBGs when it is highly/perfectly ordered, i.e., close to or at the densest packing fraction, there is a wide range of intermediate packing fractions where a PBG exists in the self-assembled (noisy) structures.
Moreover, at some intermediate packing densities, the ``randomness" of the self-assembled system improves the TM band gap compared to that of rods in the corresponding perfect lattice (provided a PBG exists). The width of the PBG is packing-density dependent, suggesting that in experiments the PBG can be adjusted through, e.g., changing the concentration of colloidal particles in suspension by adding or decreasing solvent. A comparison of rods with circular and square cross section suggests that shape can be used as another control factor to engineer the PBG of a self-assembled system. Further, we showed that by making rods hollow and optimizing the internal radius, the PBG of the TE mode can be significantly increased. Internal rod structure is but one ``dimension" (examples of other dimensions are illustrated in Ref.~\cite{Glotzer2007,Gang2011}) available to engineer particles that can produce a PBG. In all, our study suggests that by suitably choosing the packing density, particle shape, and engineering other dimensions such as particle internal structure, self-assembly can indeed be a promising method to make photonic crystals with large band gaps despite the inherent thermal noise (randomness).

\acknowledgments  
This work was supported by a grant from the Simons Foundation (256297, SCG) and used the Extreme Science and Engineering Discovery Environment (XSEDE) \cite{Towns2014}, which is supported by National Science Foundation grant number ACI-1548562; XSEDE award DMR 140129. Computational resources and services were also supported by Advanced Research Computing at the University of Michigan, Ann Arbor. D.W. also thanks the National Natural Science Foundation of China (Grant No. 11904265) for support.

\bibliography{pc_refs}

\newpage
\clearpage
\onecolumngrid
\begin{center}
\textbf{\large Supplementary Material}
\end{center}
\setcounter{equation}{0}
\setcounter{figure}{0}
\setcounter{table}{0}
\setcounter{page}{1}
\makeatletter
\renewcommand{\theequation}{S\arabic{equation}}
\renewcommand{\thefigure}{S\arabic{figure}}

\begin{figure}[h]
\centering 
\includegraphics[width=4.5in]{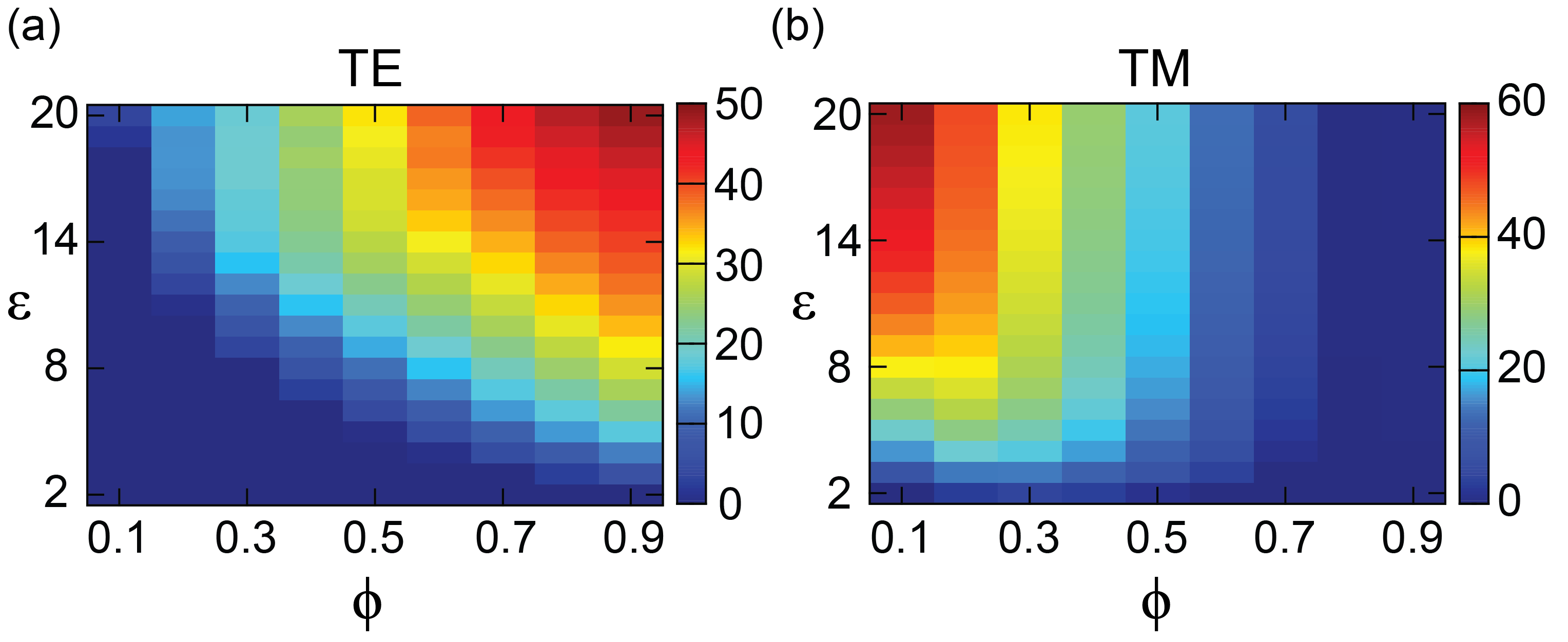} 
\caption{(Color online) Gap size as a function of $\phi$ and $\epsilon$ for hollow cylinders with optimal internal radius in a perfect triangular lattice.  }   
\label{disk_pef_opt}
\end{figure}

Fig.~\ref{disk_pef_opt} plots the band gap for hollow rods of circular cross section in a perfect triangular lattice when the inner radius is optimized. This system has much larger TE mode gap size compared to solid rods while TM gap size shows no obvious difference (see Fig.~\ref{disks}(b-c) in the main text).

\begin{figure}[h]
\centering 
\includegraphics[width=4.5in]{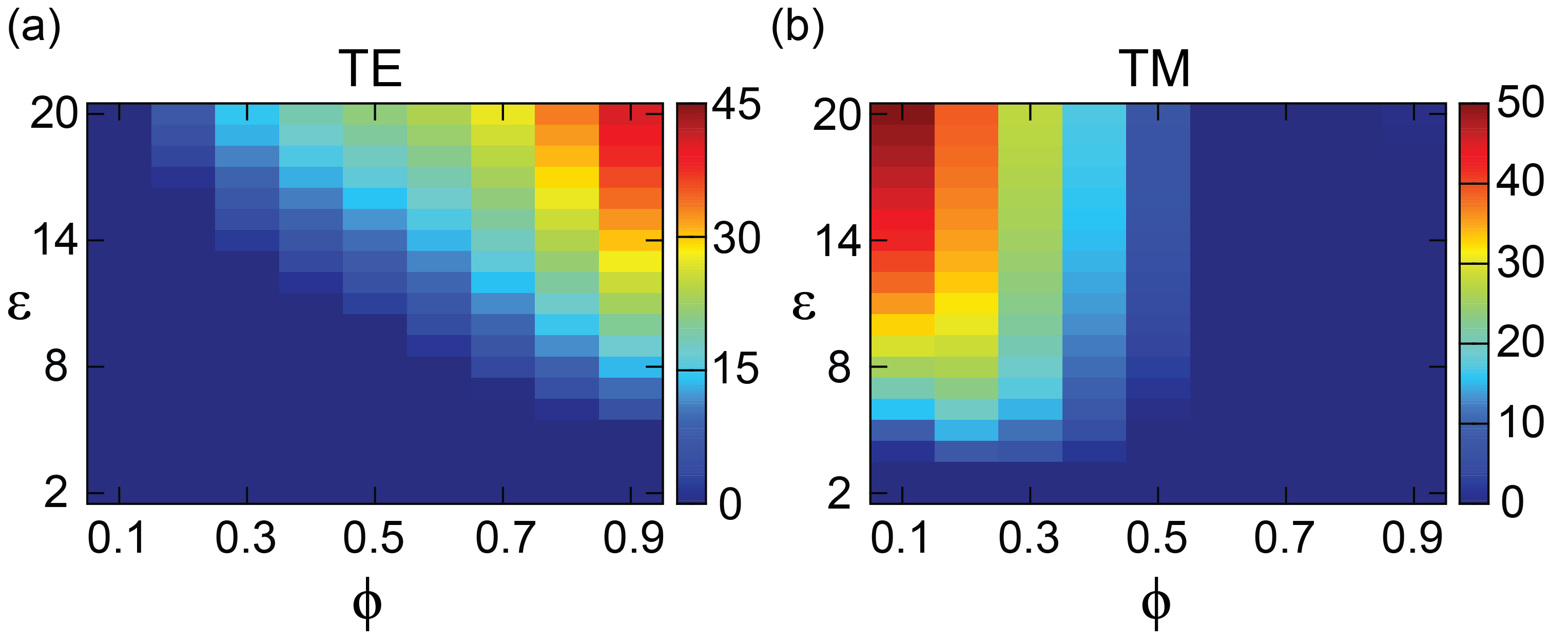} 
\caption{(Color online) Gap size as a function of $\phi$ and $\epsilon$ for rods with square cross section with optimal internal size in a perfect square lattice. }   
\label{square_pef_opt}
\end{figure}

Fig.~\ref{square_pef_opt} plots the band gap for hollow rods of square cross section in a perfect square lattice when the inner square size is optimized. Similar to the hollow circular rods, this system has much larger TE mode gap size while TM gap size shows no obvious difference, compared to its solid counterpart (Fig.~\ref{squares}(a-b)).

\begin{figure}[h]
\centering 
\includegraphics[width=5.5in]{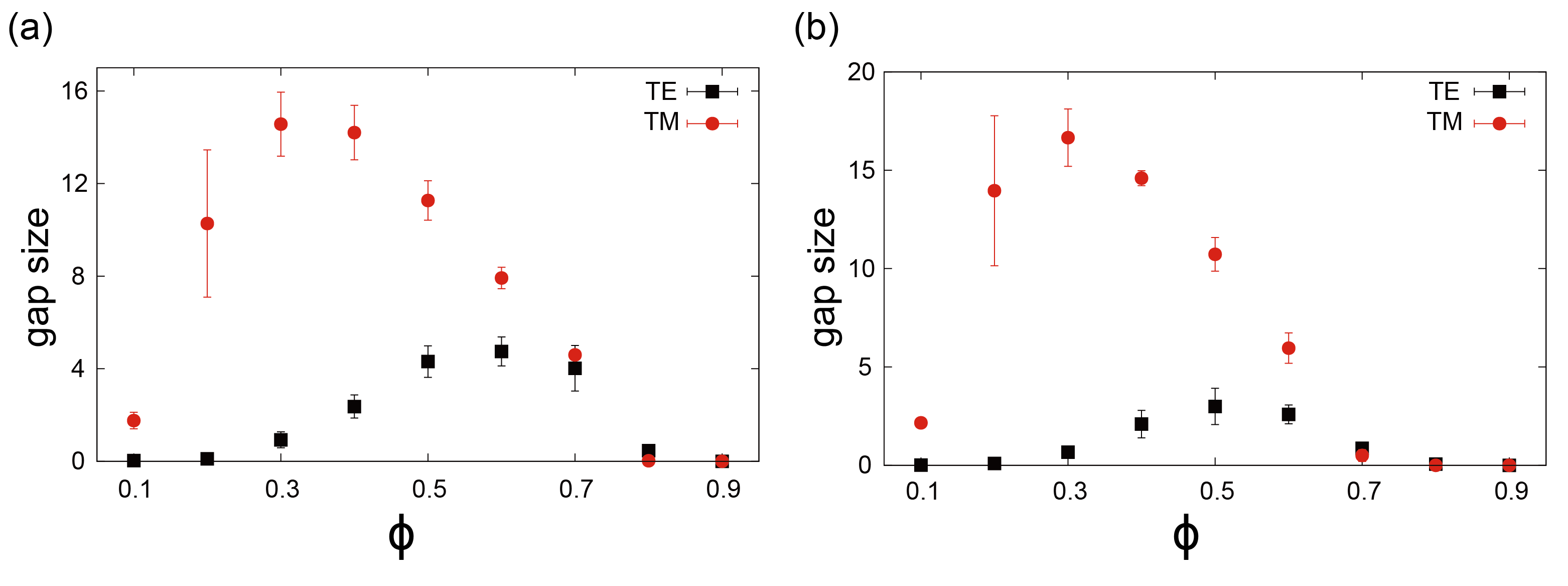} 
\caption{(Color online) Gap size as a function of $\phi$ at $\epsilon=20$ for rods with circular cross section (a) and square cross section (b). Error bars indicate the standard deviation of five independent runs.} 
\label{solid_variation}
\end{figure}

\begin{figure}[h]
\centering 
\includegraphics[width=5.5in]{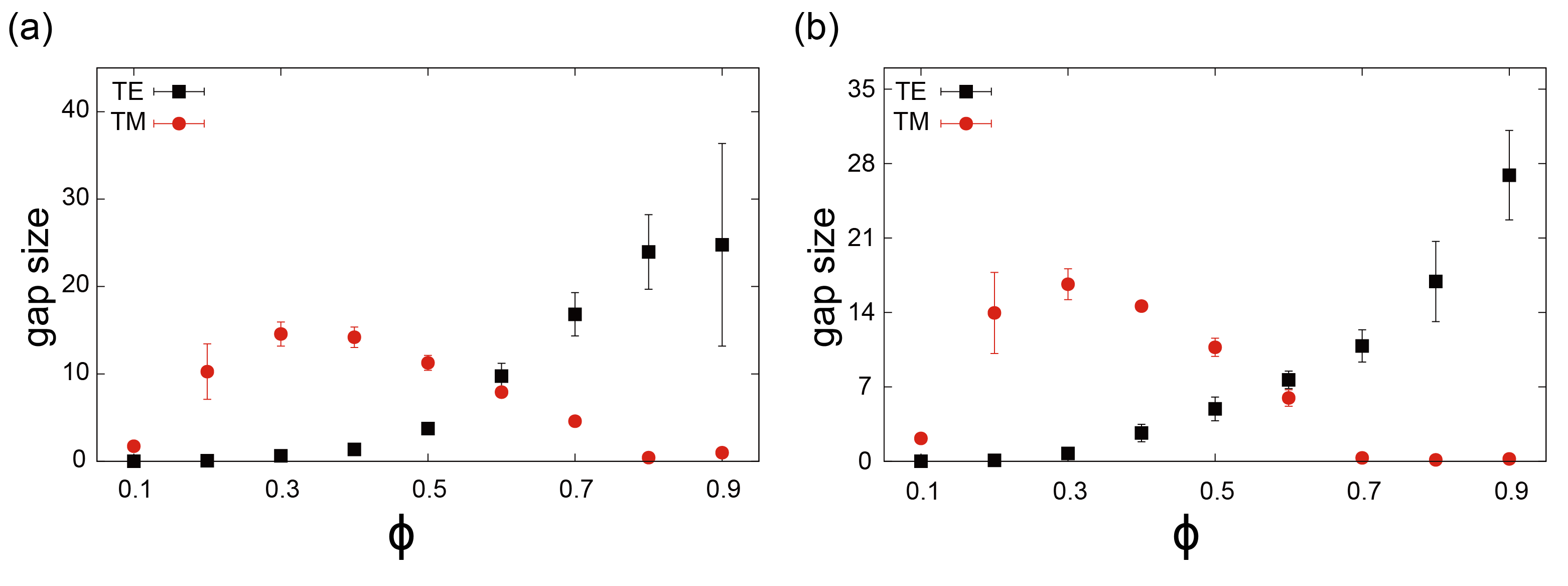} 
\caption{(Color online) Gap size as a function of $\phi$ at $\epsilon=20$ for rods of annular cross section (a) and hollow square cross section (b). Error bars indicate the standard deviation of five independent runs.} 
\label{hollow_variation}
\end{figure}

The variation of the results of the five independent runs is not very large. As an example, we plot the standard deviation at $\epsilon=20$, which has the largest PBGs. Fig.~\ref{solid_variation} refers to the solid rods (Fig.~\ref{disks}(e-f) and Fig.~\ref{squares}(c-d) in the main text) and Fig.~\ref{hollow_variation} refers to the hollow rods (Fig.~\ref{hollow}(c-d) and (g-h) in the main text).

\end{document}